      [1999/12/01 v1.4c Il Nuovo Cimento]
\documentclass{cimento}
\usepackage{graphicx}  
\usepackage{epsfig}
\title{Solar Oblateness from Archimedes to Dicke}
\author{Costantino Sigismondi\from{ins:x} and Pietro Oliva\from{ins:y}}
\instlist{\inst{ins:y}University of Rome La Sapienza \inst{ins:x} ICRA, 
International Center for Relativistic Astrophysics, and University of 
Rome La Sapienza, Piazzale Aldo Moro 5 00185 Roma, Italy.}

\PACSes{\PACSit{95.10.Gi} {Eclipses, transits, and occultations -}{95.30.Sf - Relativity and gravitation -}{95.55.Ev - Solar instruments}}

\begin{document}

\maketitle

\begin{abstract}

The non-spherical shape of the Sun has been invoked to explain the 
anomalous precession of Mercury.
A brief history of some methods for measuring solar diameter is 
presented.
Archimedes was the first to give upper and lower values for solar 
diameter in third century before Christ; we also show the method of total eclipses, 
used after Halley's observative campaign of 1715 eclipse; the variant of 
partial eclipses useful to measure different chords of the solar disk; 
the method of Dicke which correlates oblateness with luminous excess in 
the equatorial zone.
 
\end{abstract}

\section{Archimedes}

\begin{figure}
\centerline{\epsfxsize=4.1in\epsfbox{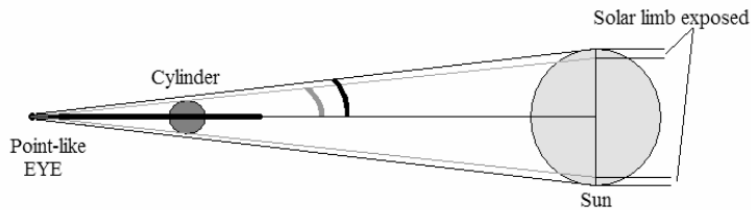}}      
\caption{Geometry of angular solar diameter measurement in case of point-like eye. 
Gray angle is the angular lower diameter, black angle is the upper limit. The ruler is the black horizontal line.}
\end{figure}

\begin{figure}
\centerline{\epsfxsize=4.1in\epsfbox{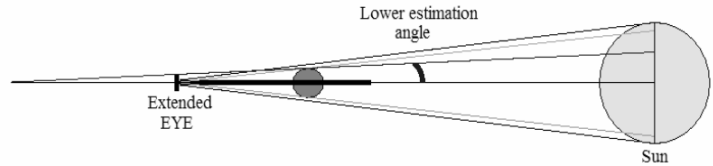}}      
\caption{Geometry of angular solar diameter measurement in case of real eye. A real eye in the same position of the point-like eye sees a major portion of solar surface. In order to see only the limb it is required to approach more 
the eye to the cylinder.}
\end{figure}

\begin{figure}
\centerline{\epsfxsize=4.1in\epsfbox{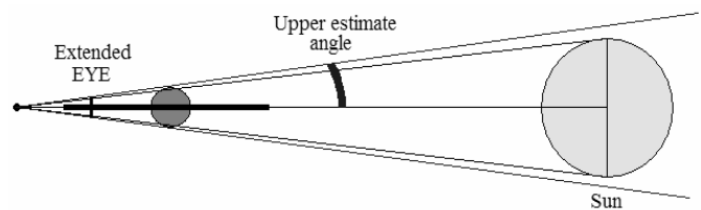}}      
\caption{Real eye geometry: the true solar angular diameter is obtained leading tangents from the eye border to the cylinder and up to the solar limb. Evaluating the solar angular diameter from the center of the eye yields an overestimate of it, as Archimedes pointed out in the $Arenarius$ or $Sand ~Reckoner$.}
\end{figure}

Archimedes of Syracuse \cite{ref:archimedes} gave out an evaluation of 
the angle subtended by the Sun with the vertex on the observer's eye. He 
knew that a perfect determination is not possible due to the 
observer bias and systematic errors, so he proposed to find out the 
upper and lower angles allowed by his method: in this way we have a 
range within solar angular diameter lies. 
To do this we must have a big ruler (a strip of wood more likely) 
mounted on a basement in a place such that is possible to see the dawn 
(Syracuse in Sicily has an eastern free sea horizon). 
When the Sun rises up and it is still near the horizon, so that is 
possible to aim at it, we point the ruler toward Sun; suddenly we put on the 
ruler a rounded cylinder such that, if we look from one edge of the 
ruler pointed at the Sun, the Sun is hidden behind that. Afterwards we 
need to shift the cylinder on the ruler till we see a very little Sun limb 
from the edges of the cylinder. Now we block the cylinder in this 
position. Let's find now, the lower evaluation angle.
 Archimedes argued: if the observer's eye was point-like (fig.1), then 
the angle identified by the lines drawn from eye's site such to be 
tangent to cylinder's edges (the gray angle in fig.1), would be smaller 
than the angle made by the lines from the eye to the edges of the Sun 
(the black angle in fig.1). This is because we set the cylinder such that 
to see a portion of the Sun. 
Indeed the observer's eye is not point-like but extended: we are 
overestimating the lower limit angle (see captions in fig. 2 and 3). 
This lead to the situation described in 
fig. 2 where we put, instead of the eye, a rounded surface with similar 
dimensions to those of the real eye. It is possible to find an appropriate 
rounded surface by taking two thin cylinders coloured black and white. 
Then, we need to put the black one close to the eye while the white one 
is placed more far in the line of sight. The right size of the surface is 
the one such that the black shape perfectly hides the white one and vice 
versa. In this way we're sure to take a rounded surface not smaller 
than the eye.
Now we need to find the upper limit angle: this is easier, we just need 
to shift the cylinder on the ruler until the Sun completely disappears 
and we block it. The angle identified by the line from the eye's border 
tangent to the cylinder is surely bigger (or equal at least but not smaller) 
than the true one because now we can't see the Sun no more (fig.3). 
Quantitatively, Archimedes measured the lower limit  1/200 times the 
squared angle and that the upper limit 1/164 times the squared angle, i.e. 
$27'\le\Theta_{\odot}\le 32'55"$. With such accuracy no oblateness was 
detectable; unless the apparent one due to atmospheric refraction, 
quantitatively studied by Tycho Brahe in sixteenth century. 

\section{Measuring solar diameter with eclipses}

The method of measuring the solar diameter with eclipses
exploits the same principle of that one of the transits of Mercury:
to recover the solar disk from more 
than three points. Those points correspond to
the external (first and fourth) or internal (second and third, 
not present in a partial eclipse) 
contacts of the Moon or Mercury with 
the solar disk, as seen by different observers
at different locations on the Earth (see figure 4).

The observations of the instants of 
totality (second and third contact) are not affected by atmospheric 
seeing, because of the sudden change of the overall luminosity. 
Conversely the determination of the instants of the external 
contacts during a partial eclipse (and all contacts during a transit), 
until the 
possibility of using CCD cameras or adaptive-optics techniques, was 
heavily affected by the seeing and the resolution of the telescope.

\subsection{Historical data}

The effect due to the telescope resolution matching with the seeing 
conditions, is evident from the 
data retrieved in the 1925 total eclipse at the 
Chamberlin Observatory in Denver, Colorado [Table 1 from Howe 
(1925)\cite{ref:Howe}].

   \begin{table}
      \caption[]{Last Contact timings of Total Eclipse of January 24, 
1925}
     $$ 
         \begin{array}{p{0.5\linewidth}ll}
            \hline
            \noalign{\smallskip}
            Aperture (inches) &Magnification ~~ power ~~~~~& Time 
^{\mathrm{a}} \\
            \noalign{\smallskip}
            \hline
           
           2".75    & 33    & h:7~46^m~55.6^s   \\
           3".4     & 80      &   h:7~47^m~02.4^s \\
           5".0     & 55      &   h:7~46^m~47.3^s  \\
           6".0     & 60     &    h:7~46^m~59.9^s  \\ 
           20".0   & 175   &    h:7~46^m~59.5^s \\
 
            \noalign{\smallskip}
            \hline
         \end{array}
  $$
\begin{list}{}{}
\item[$^{\mathrm{a}}$] The first three of these times were noted by the 
correspondent observers 
with stop watches; the last two chronographically.
\end{list}
   \end{table}

From this dataset the eclipse has lasted more when observed with larger 
instruments, 
exception made for the second data.

\subsection{New perspectives}

We propose to monitor the external contacts of a partial eclipse
with CCD cameras whose frame rate
$\Delta t \sim ~0.01~$s is below the timescale of 
atmospheric seeing. Many solar photons can be 
gathered even with a semi-professional telescope 
of diameter $d \ge 0.2 ~$m with a bandpass filter. 
CCD frames can fix the instantaneous wavefront 
path. The presence of the lunar limb helps to evaluate the
instantaneous point spread function for
reconstructing the unperturbed wavefront according 
to current image-restoration techniques (Sanchez-Cuberes et al., 
2000\cite{ref:sc}).

The identification of the lunar limb features and the solar limb 
near the contact event, in each image, and the absolute 
timing of each frame with WWV radio stations, will allow to know 
precisely the
lunar feature and the time of the contact's event. 

\begin{figure}
\centerline{\epsfxsize=4.1in\epsfbox{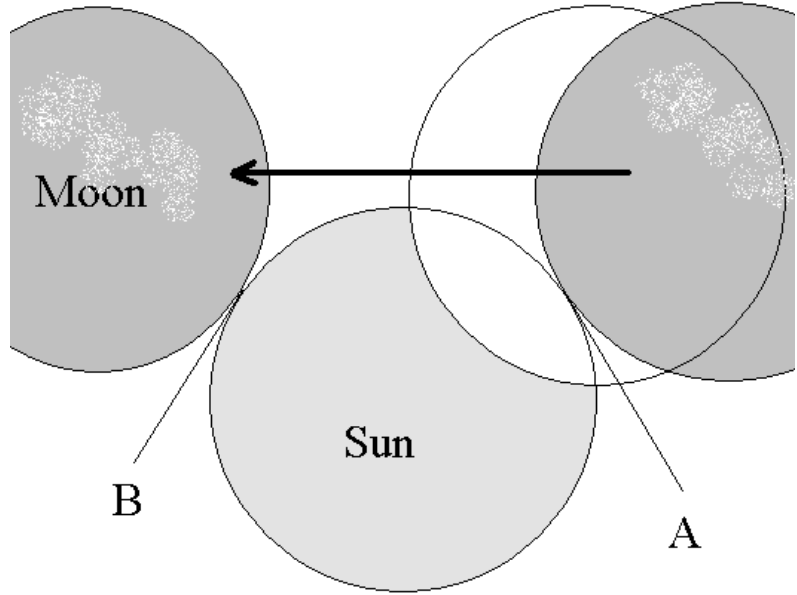}}      
\caption{In a partial eclipse we have only external contacts between solar and lunar limbs: (A) first and (B) last. Total and annular eclipses have also internal contacts which delimitate the phases of totality or annularity of the eclipse.}
\end{figure}

If our partial eclipse method can be made to succede in practice, it 
will open up more possibilities for measuring the solar diameter, 
especially because partial eclipses can be seen
from large observatories relatively often. 

\subsection{Image-restoration techniques}

The combination between finite resolution of the telescopes, 
atmosphere's turbulence and stray lights from other regions of the 
solar disk 
(both due to scattering in the Earth atmosphere and by optical 
surfaces)
can be quantitatively studied during a partial eclipse.
In fact the degradation effects made by the imaging system 
(atmosphere+telescope)
are to be considered at the exact instant of the exposure, and when 
the Moon's limb crosses the solar disk, it serves as a reference object
to estimate the amplitudes of the instantaneous optical transfer 
function.

Once known that function, there are several methods developed to 
compensate
for the atmospheric effects (Sanchez-Cuberes et al., 2000).

For example, with the 50 cm Swedish Vacuum Solar Telescope SVST at La 
Palma 
L$\ddot{o}$fdhal et al. (1997)\cite{ref:loef}
used phase-diversity speckle restoration technique to study the 
evolution of 
 bright points (0.2 arcsec of apparent dimension). 
The application of phase diversity restoration technique allows to 
reach the 
limits imposed by the diffraction to the instrument and can help 
adaptive optics
to improve them (Criscuoli et al. 2001 \cite{ref:cri}).

Recently Sanchez-Cuberes et al. (2000)\cite{ref:sc} have studied at 
high resolution
(0.53 arcsec at the solar limb, and better in other regions) 
solar granulation features from CCD images of $13~ms$ 
of exposure taken  with the SVST during the eclipse of May 10, 1994.
Their idea was to match the lunar limb present in each frame to the 
numerical 
simulation of the eclipse geometry, having included the lunar limb 
topography as
given by the Watts' profiles (Watts, 1963\cite{ref:watts}).

Although past efforts to determine the solar diameter using 
observations of partial 
solar eclipses have failed 
due to atmospheric seeing, the possibility 
to restore video CCD images can succeed in the goal of
determining with a great accuracy, for each observer whose position 
is known within $10~m$ of accuracy:

\begin{itemize}
\item{the features of the lunar limb which firstly `hits` the solar 
limb, 
and that one which is the last; with an accuracy of 0.2 arcseconds}.
\item{the instant of the external contacts of
the actual lunar limb with the solar limb, with an accuracy of 
$0.01~s$}. 
\end{itemize}

\section{Solar diameter measurements using total eclipses and transits}

The method we present here is to be compared to the 
determination of the North-South diameter of the Sun 
(which is the polar one only when $P_0=0^o$, at apsides) from the 
analysis of 
total solar eclispes observed at the edges in order to recover secular 
variations in the solar diameter (Dunham and Dunham, 
1973\cite{ref:dunham}; Fiala et al. 1994\cite{ref:fiala})
and to other determinations of the solar 
diameter based upon the observations of meridian 
transits of the Sun (see e. g. Winlock, 1853\cite{ref:winlock} and 
Ribes et al., 1988\cite{ref:ribes} 
accounting on the observations made by Picard in seventeenth century) 
or of the transits of Mercury 
across the photosphere (see Parkinson et al., 1980\cite{ref:parkinson} 
and 
Maunder and Moore, 2000\cite{ref:mm} for a complete historical review).

\subsection{Total eclipses from centerline}

Totality occurs when the solar limb disappears behind the last valley 
of the eastern lunar limb and ends when the Sun reappears 
from another depression of the western lunar limb.

A source of 
error in the evaluation of solar diameter 
arises from the knowledge of Moon's 
limb features. There are about $0.2$ arcsec of uncertainty 
in Watts' tables (1963)\cite{ref:watts}, as determined from pairs of 
photoelectrically 
timed occultations (Van Flandern, 1970\cite{ref:vanflandern}; Morrison 
and Appleby, 1981\cite{ref:morrison}). 

Therefore if one relys on Watts' profile the best determinations with 
total solar 
eclipses can not reach an accuracy better than 
$0.2$ arcsec. 
But the accuracy on the evaluation of the solar diameter can be 
considerably improved by measuring the times of dozens of Baily's 
beads phenomena, involving a similar number of Watts' points, thereby 
decreasing the error statistically.  

\subsection{Total eclipses from edges}

Even better is to make measurements 
relative to the same polar lunar valley bottoms at similar latitude 
librations, possible since all solar eclipses occur on the ecliptic 
with negligible latitude librations.
It means to observe the total eclipse near the edges (Dunham and 
Dunham, 1973\cite{ref:dunham}).

Moreover, it is possible to exploit also situations of same 
longitudinal libration angle.

The eclipses of 1925 and 1979 (after three complete Saros cycles, 
an $\sl `Exeligmos'$ 54 years and 34 days) where also exactly at the 
same 
longitudinal libration angle: their comparison (Sofia et al., 
1983\cite{ref:sofia1})
removes the uncertainty 
on the measured variations of the solar diameter due to
the Watts' errors almost entirely.
 
With current lunar profile knowledge, then total and annular 
eclipses are better for determining the solar diameter, because they
can produce polar Baily's beads when observed at the edges of their 
totaly
(annularity) path.   

An error of $10~m$ in the determination of the edges 
of the band of totality gives about $0.006$ arcsec of 
uncertainty in the evaluation of solar diameter. 

Regarding the timing of the beads events, 
the solar intensity goes to almost zero very 
quickly, then atmospheric seeing errors are more directly eliminated.  

\subsection{Transits}

The transits of Mercury of November 15, 1999, was a `grazing' 
transit (Westfall, 1999\cite{ref:westfall}), not useful for an accurate 
measure of 
the solar diameter, because it did not allow to sample points of the 
solar 
disk enough spaced between them.
The previous transit occurred in 1985 well 
before adaptive optics techniques and the large 
diffusion of CCD cameras. 
The transits of Mercury of May, 7 2003, and 
Venus (June 8, 2004 and 2012) have to be considered also
for this pourpose.

\section{Expected accuracy with partial eclipses evaluations} 

\subsection{Positions of the observers}

An error of $10~m$ in the determination of the edges 
of the band of totality gives about $0.006$ arcsec of 
uncertainty in the evaluation of solar diameter. 

An accuracy of $10~m$ in geographical position of the observer can be 
achieved 
with about $10$ minutes of averaging GPS.

\subsection{Bandpass filter}

The observations have to be done with a filter
with waveband of $6300 \pm800 \AA$, in order to have
data always comparable between them, and in the same
waveband of Solar Disk Sextant (SDS, see last paragraph).

\subsection{Duration of the imaging of the external contacts}

The eclipse magnitude $m$ is the fraction of the Sun's 
diameter obscured by the Moon.

The relative 
velocity of the Moon's limb over the Sun's 
photosphere is about $v=0.5$ arcsec per second, along the centerline 
of a total eclipse. 
For a partial eclipse the velocity of penetration of the dark figure of 
the Moon
(perpendicularly to the solar radius) is $v\sim0.5\cdot(1-m)~arcsec/s$, 
then
for having about 1 arcminute of Moon already in the solar disk it is 
necessary to continue to take images for $\Delta~t=120/(1-m)~s$ after 
the first contact and before the last contact.

The instants $t_1$ and $t_4$ of the external contacts can be 
determined with an accuracy better than the frame rate $\Delta t \le 
10^{-2}$s. 
In fact $t_1$ and $t_4$ can be deduced by interpolating 
the motion of the rigid Moon's profile, which 
becomes better defined as the 
eclipse progresses. 

In this way each observer (2 at least are needed) can fix two 
points on the Moon's limbs 
and two instants for the contacts. 

\subsection{Expected accuracy in the solar diameter measurements}

The accuracy of the determination of the lunar features producing
the external contacts for a given observer
is therefore limited by the Watts' profiles errors (0.2 arcsec).

Two observers enough distant (500 to 1000 Km in latitude for a 
East-West path of the eclipse) allow to have 4 points and 4 times 
for recovering the apparent
dimensions of the solar disk at the moment of the eclipse
within few hudredth of arcsecond of accuracy.

The accuracy becomes worse as the points sample a smaller part of the 
solar 
circle. The following table shows how the error on the determination 
of the solar diameter changes from having three 
points within 60 degrees to 240 degrees.

 That accuracy can allow the detection of the oblateness of the Sun. 
Therefore more than three 
observers can allow improvement of the detection of the shape of the 
Sun by minimizing the residuals of the best fitting ellipse.

\section{Perspectives on eclipse methods}

We have proposed an accurate measurement of the solar diameter during 
partial solar eclipses. This method is the natural extension of the 
method of measuring the solar diameter during total eclipses. It exploits 
modern techniques of image processing and fast CCD video records to 
overcome the problems arising from atmospheric turbulence.
With this method professional and semi-professional observatories can 
be involved in such a measurements, much more often than in total 
eclipses.

Moreover this method can be used for obtaining data useful for 
the absolute calibration of measurements by 
instruments that are 
balloon-borne (Sofia et al., 1994;\cite{ref:sofia2} 
1996\cite{ref:sofia2}) or satellite-borne (Dam\'e et al., 1999\cite{ref:dame}) with a 
precision of 
$\Delta D \le 40$ $10^{-3}$arcsec. 

It is also to note that from the first to the fourth contact of 
eclipses
there are about two hours. 
The apparent solar diameter changes with a maximum hourly 
rate up to $25$ $10^{-3}$arcsec/hr due to the orbital motion of 
the Earth; this effect is strongly reduced around the 
apsides on July $4^{th}$ and January $4^{th}$, $\le 2$ 
$10^{-3}$arcsec/hr and this is a favourable case for eclipses in December-January 
or June-July.

In the future Watts'
tables can be substitued by the upcoming (2004)
data of the Selene Japanese spacecraft\cite{ref:selene}, and the 
systematic 
errors arising from them will be avoided. 

\section{Secular variations of solar diameter}

It was the 3rd of May 1715 when solar eclipse was observed in England 
from both edges of the paths of totality. Following Dunham and Dunham 
method \cite{ref:dunham} it is possible to extract solar radius information 
by determining the edges of the path of the totality.
Unluckily there are elements of uncertainty on the effective positions 
of the observers on the edges \cite{ref:dunham2} and this causes a 
remarkable error on the radius determination using 1715's eclipse data.
Another eclipse on January 24, 1925 was very accurately observed by more than 
100 employees of the Affiliated Electric Companies of  NY City and many 
other advanced amateurs in response to the campaign led by E.W. Brown 
and a detailed study was made after the observation \cite{ref:brown1}. 
Sofia, Fiala et al. \cite{ref:sofia1} used Brown's data and found a 
correction of ($0.21 \pm 0.08$) arcsec. for the standard solar radius value 
of 959.63 arc sec at a distance of 1UA, for 1925 eclipse.
Analyses of the eclipse in Australia in 1976 and of the eclipse in 
North America in 1979, were made by Sofia, Fiala et al. \cite{ref:dunham2} 
but no appreciable changes in the solar radius were found between those 
two eclipses. However, the solar radius determined for 1715 was found 
to be ($0.34  \pm 0.2$) arcsec larger than 1979 value.
On the other hand, Sofia, Fiala, Dunham and Dunham \cite{ref:sofia1} found that between 
the 1925 and the 1979 eclipses, the solar radius decreased by 0.5 arcsec 
but the solar size between 1925 and 1715 did not significantly 
changed. Therefore they concluded that the solar radius changes are not 
secular. 
Eddy and Boornazian \cite{ref:eddy} in the same year reported results 
over observations made between 1836 and 1953 at the Royal Greenwich 
Observatory. They found a secular decrease trend in the horizontal solar 
diameter amounting to more than 2 arc sec/century while the solar 
vertical diameter seemed to change with about half of this rate. 
With the same data Sofia et al. \cite{ref:sofia1} had found out that any 
secular changes in the solar diameter in the past century, could not have 
exceeded 0.25 arc sec. The disagreement between the results of different groups
depends on the different data selection criteria and on different solar and lunar 
ephemerides adopted, as it is shown in \cite{ref:fiala} for the analyses of the annular eclipse of May 30, 1984.
Another measurement of the solar radius, independent on lunar ephemerides, was made by Shapiro 
\cite{ref:shap} who analyzed data from 23 transits of Mercury between 1736 
and 1973. His conclusion was that any secular solar radius decrease was 
below 0.15 arc sec/century. This method has been criticized for the black drop effect which affects the 
exact determination of the instans of internal contacts, first pointed out by Captain Cook during Venus' transit of 1769.

\section{The method of Dicke for measuring solar oblateness}

Around 1961, R. H. Dicke and others\cite{ref:dicke} tried to point out 
the possible effects due to existence of a scalar field in the 
framework of Einstein's General Relativity. The presence of such a scalar field 
would have important cosmological effects. The gravitational deflection 
of light and the relativistic advancement of planetary perihelia are two 
effects that could have been influenced by a scalar 
field: with respect to classical General Relativity both effects were expected to be about $10\%$ less in the case 
the scalar field would be present.
For this reason Dicke showed that the advancement of the line of apsides of Mercury 
was not to be considered as a good test for General Relativity, which was 
believed before, because of the entanglement of its causes (scalar field and classical General Relativity)
\cite{ref:dicke2}. 
A small solar oblateness ($\Delta R/R \sim 5\cdot10^{-5}$) 
caused by internal rotation in the Sun would cause the $10\%$ effect of 
perihelion advancement without invoking any relativistic effect.
It was clear that until such oblateness could be excluded or confirmed 
from observational data, the interpretation of the advancement of 
Mercury's line of apsides would was ambiguous.
The Einstein relativistic motion of the longitude of the perihelion is

     (1)  $\dot{\pi}=\frac{1}{Tac^2(1-e^2)}$

where $a$ is the planetary semimajor axis, $e$ is the eccentricity and $T$ is 
the period; on the other hand we have the rotation of the perihelion 
due to an oblate Sun which is

     (2)  $\dot{\pi}=\frac{\Delta}{Tac^2(1-e^2)^2}$

where $\Delta$ is the ratio between (solar equatorial radius - polar 
radius) and (mean radius). The scalar-tensor theory of gravitation could 
have been brought in agreement with observational data, if the Sun 
possessed a small oblateness and a mass quadrupole moment.

\begin{figure}
\centerline{\epsfxsize=4.1in\epsfbox{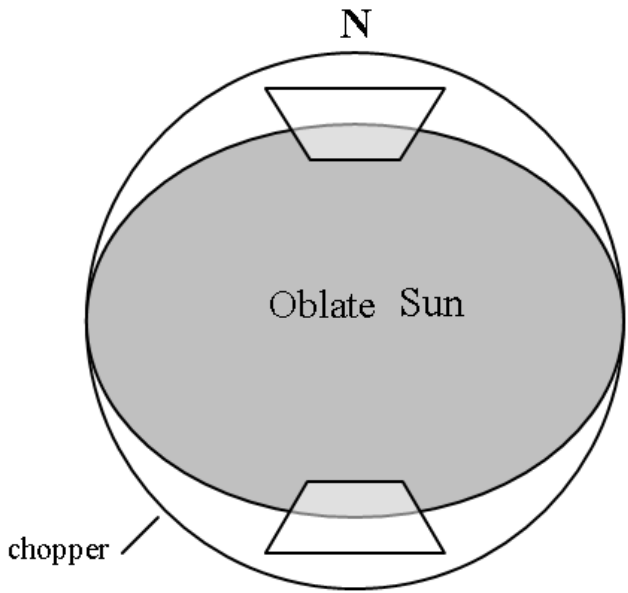}}      
\caption{Measurement of the flux $F$ within the chopper mask at the solar poles. 
By changing the fraction $f_l$ of exposed limb it is possible to detect if there is any $\Delta F$ between 
polar and equatorial diameters, and if it changes with $f_l$. Only if $\Delta F \not= 0$ and it is constant with $f_l$
it is a consequence of the solar oblateness, otherwise it would be a consequence of temperature gradients.}
\end{figure}

In 1966, Dicke and Goldenberg \cite{ref:dicke3} measured the difference 
in flux between the equator and polar limb of the Sun. The idea was 
simple: using a chopper with apertures made to show only a small section 
of the solar limb (see fig. 5), they measured the flux at the poles and 
at the equator of the Sun. We must consider two ipothesis:

\begin{itemize}
\item{If the temperature at the pole is equal to the one at the equator, 
finding a flux difference can only mean that there is an oblateness such 
that the radius at the pole $R_p$ and the radius at the equator $R_e$, 
differs from a quantity $\Delta R$. Then the flux difference $\Delta F$ 
should be constant if we change the exposed limb by changing the 
chopper's aperture.}

\item{On the contrary, if there is no oblateness but we still have a flux 
difference $\Delta F$, means that there is a temperature gradient. Then 
$\Delta F$ should be proportional to the amount of exposed limb.} 
\end{itemize}

Dicke and Goldenberg found that $\Delta F$ remained about constant so 
that the Sun should have a small oblateness. 
The theory of the measurement is well explained in \cite{ref:dicke4}, § 
II. The instrument used and the measurement procedure is also explained 
in \cite{ref:dicke4}, § IV, V, VI and VII.
Mainly both the instrument and the measuring procedure were designed to 
eliminate systematic errors. Dicke found for $\Delta R$ to be $\Delta R = 43.3 \pm 3.3$ $10^{-3}$arcsec.     

The oblateness of  $\Delta R/R = (4.51 \pm 0.34)\cdot 10^{-5}$ implies 
a quadrupole moment of  $J = (2.47 \pm 0.23 )\cdot 10^{-5}$ 
\cite{ref:dicke5}.

At the end of his analysis, Dicke found that new independent 
measurements of the solar oblateness were needed, to make comparison between 
data taken with different faculae activity on the Sun.
In 1975, new observations were made by Hill and Stebbins 
\cite{ref:hill}. They considered a complication raised in Hill's work 
\cite{ref:hill2} of a time varying excess of equatorial brightness due to sunspots and 
faculae. It is clear that to measure the difference between the 
polar radius and the equatorial radius, we must first be sure on which 
point to take as equatorial solar edge.
The point is to give out a consistent definition of the solar limb. 
This can be done by using a proper limb darkening function.
Hill et al. demonstrated that the excess brightness can be easily 
monitored by using a proper analytic definition of the solar edge, using the 
FFTD \cite{ref:hill3}.
It was pointed out that the main problem in this kind of measurements 
is identifying some point on the limb darkening curve as the solar edge.
It is clear that if more points on the darkening curve can be taken as 
solar edge, many different definition of solar radii can be gave and 
many different measures of solar oblateness done. The differences between 
these values will contain information about the shapes of the limb 
profiles.
The value obtained from Hill for the intrinsic visual oblateness is 
$(18.4 \pm 12.5)\cdot 10^{-3}$ arcsec which is obviously in conflict with 
the value of Dicke-Goldenberg.
In this confused situation another group decided to construct an 
instrument to measure long term changes: the Solar Diameter Monitor (SDM) at 
The High Altitude Observatory \cite{ref:g15}. Their purpose was to 
determine which kind of solar diameter variation was taking place, if any, 
within a reasonable period of time (3-5 years). The 
SDM began operation in Aug. 1981. An accurate discussion on the 
measured duration of solar meridian transit during six years between 1981 and 
1987 is made in \cite{ref:brown} where Brown and Christensen-Dalsgaard 
adopted adjustments to the modified IAU value for the astronomical unit 
(value of $1.4959787066 \cdot 10^5$ Mm, US Naval Observatory, 1997) to 
take into account for the mean displacements between the telescope's 
noontime location and the Earth's centre. They also corrected for the 
displacements of the Sun's centre relative to the barycentre of the 
Earth-Sun system.
They found the solar radius to be $R_{\odot} = (695.508 \pm 0.026)$ Mm
which is about 0.5 Mm smaller than the Allen (1973) value of 695.99 Mm.
Moreover, Brown and Christensen-Dalsgaard found no significant 
variations in the solar diameter during their observational period: their 
annual averages for the years 1981-1987 all agree within $\pm 0.037$ Mm.
Toulmonde \cite{ref:toul} discussed 
about 71000 measurements regarding almost 300 years of 
data: he did not find
evidence of any secular variation in his data.

\section{Solar Disk Sextant measurements}

Further attempts to measure the solar oblateness have been made
with the Solar Disk Sextant (SDS) which is an instrument made to 
monitor the size and shape of the Sun. The principle of the instrument 
is well described in Sofia, Maier and Twigg work \cite{ref:sofiamaier}.
Basically a prism whit an opening angle very stable along the years is posed in front 
of the objective of a telescope, and it produces tweo images of the Sun at focal plane.
The distance between the center of those images is depending on the focal lenght of the telescope, while
the gap between the two limbs depends on the angular diameter of the Sun. 
The same idea is exploited in using two pinholes instead 
of one and has been proposed for simpler prototypes of SDS\cite{ref:sigi2pin}, 
whose images are unaffected by optical distortions. 
Considered that the solar radius changes until now reported are to be of 
the order of 1 arc sec per century, the SDS instrumental accuracy was 
asked to keep calibrated on 0.01 arc sec/year and a stability of 0.003 
arc sec/year was reached.
The really good feature of the SDS consists in the fact that the 
instrument accuracy requirements are for relative rather than absolute values 
of the radius which led to a solar edge point detection 
accurate to 1/10 pixel on the instrument. With statistical methods one can 
have a further reduction of a 10 factor.
The SDS early version was developed to be carried into space 
during Space Shuttle flights, but unlikely the Challenger accident took place. 
This led to the needs to change strategy avoiding important delays. SDS was 
mounted on a system for ground based observations but it was soon clear 
that no valuable scientific data could be obtained from ground because 
the atmosphere's influence. So the SDS was mounted on a stratospheric 
balloon and it measured solar oblateness\cite{ref:sofia3}.
A complete analysis of his 4 flights data (1992, 1994, 1995 and 1996) 
is still in progress.

\acknowledgments
Costantino Sigismondi thanks Drs. Terry Girard, David Dunham and Elliot 
Horch, 
who encouraged him, during his scholarship at Yale University 
(2000-2002), to pursue the idea of partial eclipse measurements of solar 
diameter.


\begin{thebibliography}{0}

\bibitem{ref:archimedes} \BY{Archimedes} \TITLE{Psammites, The Sand 
Reckoner}, Italian edition in {\sl{Classici della Scienza}} {\bf{19}} UTET Torino (1974) p. 443-470. 

\bibitem{ref:fiala} \BY{Fiala, A. D.; Dunham, D. W.; Sofia, S.} 
\IN{Solar Physics}{152}{1994} {97} 
 
\bibitem{ref:winlock} \BY{Winlock, J.} \IN{Astronomical 
Journal}{3}{1853}{97-103}
      
\bibitem{ref:Howe} \BY{Howe, H. A.} \IN{Popular 
Astronomy}{33}{1925}{280} 
 
\bibitem{ref:parkinson} \BY{Parkinson, J. H.; Morrison, L. V.; 
Stephenson, F. R.}\IN{Nature}{288} {1988}{548-551}
\TITLE{The constancy of the Solar diameter over the past 250 years}

\bibitem{ref:mm}\BY{Maunder, M. and P. Moore}\TITLE{Transit when a planet crosses the Sun}
{Springer-Verlag London}{2000}

\bibitem{ref:watts}  \BY{Watts,C. B.} \TITLE{The Marginal Zone of the 
Moon} {1963}
Astronomical Papers prepared for the use of the
American Ephemeris and Nautical Almanac XVII
(United States Government Printing Office, Washington)

\bibitem{ref:westfall} \BY{Westfall, J. E.} {1999}
{\rm 
http://www.lpl.arizona.edu/\~rhill/alpo/transitstuff/merc11\_99.html} 
 
\bibitem{ref:selene}\BY{Hirata, N. et al.} \TITLE{General overview of 
the lunar imager/spectrometer 
in "New Views of the Moon, Europe, Future Lunar Exploration, Science 
Objectives, and Integration of Datasets", David Heather editor, (Berlin, 
Germany)} {2002}

\bibitem{ref:sofia2} \BY{Sofia, S., W. Heaps, and L. Twigg} 
\IN{Astrophys. J.}{427}{1994}{1048} \TITLE{The Solar Diameter and Oblateness 
Measured by the Solar Disk Sextant on the 1992 September 30 Balloon Flight}

\bibitem{ref:sofia3} \BY{Sofia, S. Lydon T. J.} \IN{Physical Review 
Letters}{76}{1996}{177-179} \TITLE{A measurement of the shape 
of the solar disk: The solar quadrupole moment, the solar octopole 
moment, 
and the advance of perihelion of the planet Mercury}

\bibitem{ref:dame} \BY{Dam\'e, L. et al} \IN{Advances in Space 
Research}{24}{1999}{205-214}
\TITLE{PICARD: simultaneous measurements of the solar diameter, 
differential rotation, solar constant and their variations}
 
\bibitem{ref:loef} \BY{L$\ddot{o}$ fdhal, M. G., et al.} \TITLE{ 
Phase-diversity Restoration of two Simultaneous 70-minute Photospheric 
Sequences,
Bulletin of the American Astronomical Society, 29, Volume 29, Number 2
AAS 190th Meeting, Winston-Salem, NC, June 1997} {1997}

\bibitem{ref:cri} \BY{Criscuoli, S., J.A. Bonet, F. Berrilli, D. Del 
Moro 
and A. Egidi}
\TITLE{Phase diversity procedure in F95 for future Themis application,
Meeting - THEMIS and the New Frontiers of Solar Atmosphere Dynamics 
       Roma 19-21 March, 2001}  {2001} 

\bibitem{ref:sc} \BY{S\'anchez Cuberes, M., Bonet, J. A., V\'azquez, 
M., Wittmann, A. D.}
\IN{Astrophys. J.}{538}{2000}{940-959} \TITLE{Center-to-Limb Variation 
of Solar Granulation from Partial Eclipse Observations}

\bibitem{ref:ribes} \BY{Ribes, E., et al.} \IN{Nature} {332}{1988}{689}
\TITLE{Size of the Sun in the Seventeenth Century}

\bibitem{ref:vanflandern} \BY{Van Flandern, T.} \IN{Astronomical 
Journal}{75}{1970}{744}
\TITLE{Some Notes on the Use of the Watts Limb-Correction Charts} 

\bibitem{ref:morrison}\BY{Morrison L. V. and G. M. Appleby}\IN{MNRAS} 
{196}{1981}{1005}
\TITLE{Analysis of Lunar Occultations}

\bibitem{ref:dunham}\BY{D.W. Dunham and J.B. 
Dunham}\IN{Moon}{8}{546}{1973}

\bibitem{ref:dunham2}\BY{D.W. Dunham, S. Sofia, A.D. Fiala et 
al.}\IN{Science}{210}{1980}{1243-1245}

\bibitem{ref:brown1}\BY{Brown, E. W.}\IN{Astron. J.}{37}{1926}{9-19}

\bibitem{ref:sofia1}\BY{Sofia, Dunham \& Dunham and Fiala} \IN{Nature} 
{304}{1983}{522-526}

\bibitem{ref:eddy}\BY{J.A. Eddy and A.A. Boornazian}\IN{Bull. Am. 
Astron. Soc.} {11} {1979} {437}

\bibitem{ref:shap}\BY{Shapiro, I. I.}\IN{Bull. Am. Astron. 
Soc.}{208}{1980}{51}

\bibitem{ref:dicke}\BY{Brans, C. and Dicke, R.H.}\IN{Phys. 
Rev.}{124}{1961}{925}

\bibitem{ref:dicke2}\BY{Dicke, H.R.}\IN{Nature} {202}{anno?} {432}

\bibitem{ref:dicke3}\BY{Dicke, H.R. and Golenberg, H.M.}\IN{Phys. Rev. 
Letters}{18}{1967}{313}

\bibitem{ref:dicke4}\BY{Dicke, H.R. and Golenberg, H.M.}\IN{Astrophys. 
J. supplement series} {241}{1974} {27:131-182}

\bibitem{ref:dicke5}\BY{Dicke, H.R.}\IN{Astrophys. J.}{159}{anno?} {1}

\bibitem{ref:hill}\BY{Hill, H.A. and Stebbins, R.T.}\IN{Astrophys. 
J.}{200}{1975}{471-483}

\bibitem{ref:hill2}\BY{Hill, H.A.et al.}\IN{Phys. Rev. 
Letters}{33}{1974}{1497}

\bibitem{ref:hill3}\BY{Hill, H.A., Stebbins, R.T. and Oleson, J.R.}  
\IN{Astrophys. J.}{200}{1975}{484-498}

\bibitem{ref:g15}\BY{T.M. Brown, D.F. Elmore, L. Lacey and H. Hull} 
\IN{Applied Optics}{21}{1982}{19}

\bibitem{ref:brown}\BY{Brown, T.M. and Christensen-Dalsgaard, J.}   
\IN{Astrophys. J.}{500}{1998}{L195-L198}

\bibitem{ref:toul}\BY{Toulmonde, M.} \IN{Astron. \& 
Astrophysics}{325}{1997}{1174-1178}

\bibitem{ref:sofiamaier}\BY{Sofia, S., Maier, E. and Twigg, L.} 
\IN{Adv. Space Res.}{11}{1991}{(4)123-(4)132}

\bibitem{ref:sigi2pin} \BY{Sigismondi, C.} 
\TITLE{Measuring the angular solar diameter using two pinholes,} 
\IN{Am. J. of Physics}{70}{2002}{1157}
 
\end{thebibliography}
\end{document}